\documentclass{aa}
\usepackage{graphicx}
\usepackage{txfonts}
\usepackage{color}
\usepackage{makecell}

\usepackage{natbib}
\bibpunct{(}{)}{;}{a}{}{,} 
\newcommand{\xmm}{{\it XMM-Newton}}
\newcommand{\swift}{{\it Swift}}
\newcommand{\lumUnits}{{$\,$erg$\,$s$^{-1}$}}

\begin{document}

        \title{Long-term X-ray evolution of SDSS J134244.4+053056.1:\\
                A more than 18 year-old, long-lived IMBH-TDE candidate }

        \titlerunning{A more than 18 year-old, long-lived IMBH-TDE candidate}
        
        \author{J. S. ~He
                \inst{1} 
                \and
                L.M.~Dou\inst{1}\thanks{Corresponding author: L.M.~Dou} 
                \and
              Y.L.~Ai\inst{2, 6} 
              \and
              X.W. ~Shu\inst{3} 
              \and
              N.~Jiang\inst{4} 
             \and
             T.G.~Wang\inst{4}               
             \and
             F.B.~Zhang\inst{3} 
             \and
             R.F.~Shen\inst{5}               
             }
        \institute{Department of Astronomy, Guangzhou University, Guangzhou 510006, China; 
              \email{doulm@gzhu.edu.cn}   
              \and
               College of Engineering Physics, Shenzhen Technology University, Shenzhen 518118, China               
               \and
               Department of Physics, Anhui Normal University, Wuhu, Anhui 241000, China
               \and
                CAS Key Laboratory for Researches in Galaxies and Cosmology, Department of Astronomy, University of Science and Technology of China, Hefei, Anhui 230026, China
                \and
                School of Physics and Astronomy, Sun Yat-Sen University, Zhuhai, 519082, China  
                \and    
                Shenzhen Key Laboratory of  Ultraintense Laser and Advanced Material Technology, Shenzhen 518118, China
        }

\abstract
{ 
SDSS J134244.4+053056 is a tidal disruption event candidate with strong temporal coronal line emitters and a long fading, mid-infrared dust echo. We present detailed analyses of X-ray emission from a {\swift}/XRT observation in 2009 and the most recent {\xmm/pn} observation in 2020. The two spectra can be modeled with hard and soft components. While no significant variability is detected in the hard component above 2 keV between these two observations, the soft X-ray emission in 0.3-2 keV varies by a factor of $\sim5$. The luminosity of this soft component fades from $\sim1.8\times10^{41}$ to $\sim3.7\times10^{40}${\lumUnits} from the observation in {\swift} to that of {\xmm}, which are 8 and 19 years after the outburst occurred, respectively. The evolution of luminosity matches with the {$\it t$$^{-5/3}$}decline law well; there is a soft X-ray peak luminosity of 10$^{44}${\lumUnits} at the time of the optical flare. Furthermore, the spectra of the soft component harden slightly in the decay phase, in which the photon index $\Gamma$ varies from $4.8^{+1.2}_{-0.9}$ to $3.7\pm0.5$, although they are consistent with each other if we consider the uncertainties. Additionally, by comparing the BH mass estimate between the ${\it M-\sigma}$ correlation, the broad H$\alpha$ emission, and the fundamental plane relation of BH accretion, we find that a value of $\sim10^{5}$\,{{\it M}$_{\sun}$} is favored. If so, taking its X-ray spectral variation, luminosity evolution, and further support from theory into account, we suggest that SDSS J134244.4+053056 is a long-lived tidal disruption event candidate lasting more than 18 years with an intermediate-mass black hole. 
}


\maketitle

\section{Introduction}

A tidal disruption event (TDE), in which a star passing too close to a supermassive black hole (SMBH) in a galactic nucleus is torn apart by the tidal force of the hole, is of particular interest. As a result of a TDE, about half of the stellar debris is accreted by the SMBH, producing a luminous electromagnetic flare \citep[e.g.,][]{rees1988tidal,evans1989the}.  
The transient luminosity would follow the same time dependence as the mass fallback rate, which is generally assumed  as {$\it t$$^{-5/3}$}. For a Schwarzschild black hole (BH) with {$M_{\mathrm {BH}} >\,10^8\,$${\it M}$$_{\sun}$} the tidal disruption radius of a solar-type star  is in the event horizon and no electromagnetic flare would be observed \citep[][]{hills1975possible}. 

The first TDE candidate was discovered in the soft X-ray band by the {\it {ROSAT}} all-sky survey in the 1990s and can be well interpreted by the TDE scenario \citep{bade1996detection,komossa1999the}.  The general picture for soft X-ray TDEs is that at first the flare is  up to several 10$^{44}$\,erg\,s$^{-1}$ in soft X-ray band with an Eddington accretion rate. Then, the luminosity declines on a timescale of months to years. Finally, the hardening of X-ray spectra indicates that the disk and corona coexist on the phase with a low accretion rate \citep{komossa2015tidal,saxton2020x}. Approximately 20 soft X-ray TDEs have been discovered until now \citep[see the review in][]{saxton2020x}.

As a consequence of the rapid development of time-domain astronomy in recent years, especially large optical sky surveys, the number of TDEs has increased to $\sim100$. Most of these have been discovered in the optical band, and many of the optical TDEs have no or weak X-ray detection in their very early phases during the outbursts. However, X-ray emissions have been detected in three out of four optically discovered TDEs about four to nine years after the optical flares, and most TDEs have been suggested as bright X-ray sources visible for at least a decade \citep{jonker2020implications}. Therefore, research in the X-ray band is still one of the most important tools for identifying and studying TDEs.

In the gas-rich environment, the extreme ultraviolet (EUV) X-ray emission from TDEs can photoionize the surrounding gas, thereby producing luminous, transient high-ionization coronal emission lines \citep[e.g.,][]{komossa2008discovery,wang2011transient,wang2012extreme}. While in a dust-rich environment, high-energy X-ray photons from the TDEs are absorbed by dust and then reradiated in the infrared band \citep{lu2016infrared}. Such light echoes have been observed in the mid-infrared light curves of some TDEs \citep[][]{dou2016long,dou2017discovery,jiang2016wise,jiang2017mid,jiang2019infrared,van2016discovery} and used as an efficient method to search for TDE candidates \citep[][]{wang2018long,jiang2021}.

SDSS J134244.4+053056 (hereafter J1342) is a TDE candidate with transient extreme optical coronal emission lines at a redshift of 0.0366. Strong high ionization coronal lines appeared in a SDSS spectrum (e.g.,  [Fe {\sc X}], [Fe {\sc XI}] and [Fe {\sc XIV}]) on April 09, 2002, but disappeared in the following multiple mirror telescope (MMT) spectrum on December 26, 2011 \citep{wang2012extreme,yang2013long}. The long-fading mid-infrared emission was also detected in J1342, which was interpreted as an infrared echo from dust in the nuclear region by reprocessing the high-energy radiation of the TDE \citep{dou2016long}. Although we do not have good constraints on the date for its continuum flare, the high-ionization iron coronal lines disappeared in subsequent observations; this suggested that they must be short lived. Thus, the tidal disruption flare happened not too far from the SDSS spectroscopic observation. The peak of the flare is then roughly estimated to occur roughly one year before the SDSS optical spectral observation \citep[see the details in][]{dou2016long}. 

Soft X-ray emission was detected in J1342 with {\swift}/XRT observations in May 2009, and this source was classified as a veiled X-ray TDE \citep{auchettl2017new}. However, \citet{chilingarian2018population} argue that the detected X-ray emission was not from a TDE because the observations took place seven
years after the time of the speculated optical flare, and these authors attributed the X-ray emission to active galactic nucleus (AGN) activity. Thus, the nature of its X-ray emission is not yet clear. In this paper, with the  latest {\xmm} observation in 2020 and the radio observation with VLA in 2016, we present a detailed X-ray and radio analysis of J1342. In section\,\ref{2}, we describe the observations and data reductions. In section\,\ref{3}, we present our results of X-ray spectral fitting and discuss the observed properties in AGN and TDE scenarios, respectively. We conclude in section\,\ref{4}. Throughout this paper, we adopt a standard flat cosmology with ${\Omega_{\mathrm{M}}}$=0.3, ${\Omega_{\Lambda}}$=0.7, and a Hubble constant of H$_{0}$\,=\,70\,km\,s$^{-1}$\,Mpc$^{-1}$, which results in a luminosity distance of 161.1\,Mpc.

\section{Data reduction and analysis} \label{2}

\subsection{X-ray data reduction}

Three observations of J1342 were taken with the X-Ray Telescope \citep[XRT;][]{burrows2005swift} at the \textit{Neil Gehrels Swift Observatory} \citep[{\swift};][]{gehrels2004swift} during May 15-20 2009. We reduced the data following the standard XRT data reduction\footnote{http://swift.gsfc.nasa.gov/analysis/xrt\_swguide\_v1\_2.pdf} with the tools in HEASOFT (v6.26.1).  The  cleaned events files were reprocessed using the task XRTPIPELINE. Only observations with a ``photon counting'' mode were used. The source spectrum was extracted from a circular region with radius of $47.2''$ at the optical center with the tool XSELECT.  The background spectrum was extracted from a larger source-free region near the source. No significant variation was detected among the observations and we stacked the three observations to increase the signal-to-noise ratio (S/N). The net count rate of the target is 0.00139$\pm$0.00033, 0.00016$\pm$0.00016, and 0.00154$\pm$0.00036\,cts\,s$^{-1}$ in the 0.3-2, 2-10, and 0.3-10\,keV bands, respectively.

J1342 was observed with {\xmm} on January 8, 2020 (ObsID: 0843440201; PI: Chilingarian). We reprocessed the {\xmm} data with Science Analysis Software (SAS, v19.0)  and the latest calibration files (updated to 2020 July).  We only used data from the pn instrument of European Photon Imaging Camera (EPIC-pn)  in our analysis considering its high sensitivity and large effective area. The events file was created by `epchain'. After removing the `bad' (e.g., hot, dead, or flickering) pixels, the high flaring particle background time interval was created by applying a threshold rate of $>\,$0.5\,cts\,s$^{-1}$ with single events (PATTERN=0) in the 10-12\,keV band for EPIC-pn. This results in a net exposure time of 12.46\,ks, after removing the high flaring particle background time interval. Only single and double events (PATTERN$<$=4, FLAG=0) were used for the following analysis. The source spectrum was extracted from a circular region with radius of $20''$ centered on its optical position, and the background spectrum was extracted from a nearby source-free circular region with radius of $30''$. The net count rate is  0.0093$\pm$0.0011, $0.0016\pm0.0006$, and 0.0109$\pm$0.0012\,cts\,s$^{-1}$ in the 0.2-2, 2-10, and 0.2-10\,keV bands, respectively.

The logs of {\swift}/XRT and {\xmm}/pn observations are listed in Table\,\ref{table1}. The hardness ratio (HR) between 0.3-2 and 2-10\,keV\footnote{The hardness ratio is defined as HR = (H-S)/(H+S), where S and H are the source counts for the soft (0.3-2\,keV) and hard (2-10\,keV) bands.} is $-0.83^{+0.09}_{-0.17}$ for the stacked {\swift}/XRT observation, and $-0.65^{+0.12}_{-0.12}$ for the {\xmm}/pn observation, which is calculated by the Bayesian estimation of hardness ratios code \citep[BEHR;][]{park2006bayesian}. The values of hardness ratios indicate a slight spectral shape variation between {\swift} and {\xmm} observations, although they are consistent with each other if considering the uncertainties. The {\swift}/XRT stacking spectrum ({\swift} spectrum, hereafter) is then regrouped to have at least three counts per bin, while the {\xmm}/pn spectrum ({\xmm} spectrum, hereafter) is regrouped to least seven counts per bin to adopt  the C-statistic for spectral fitting in XSPEC (v. 12.9.1). During the spectral fitting, the uncertainties are estimated at 68\% confidence level for one interesting parameter.

\subsection{X-ray spectral analysis} \label{2.2}

\begin{figure*}
        \centering
        \includegraphics[width=0.95\textwidth]{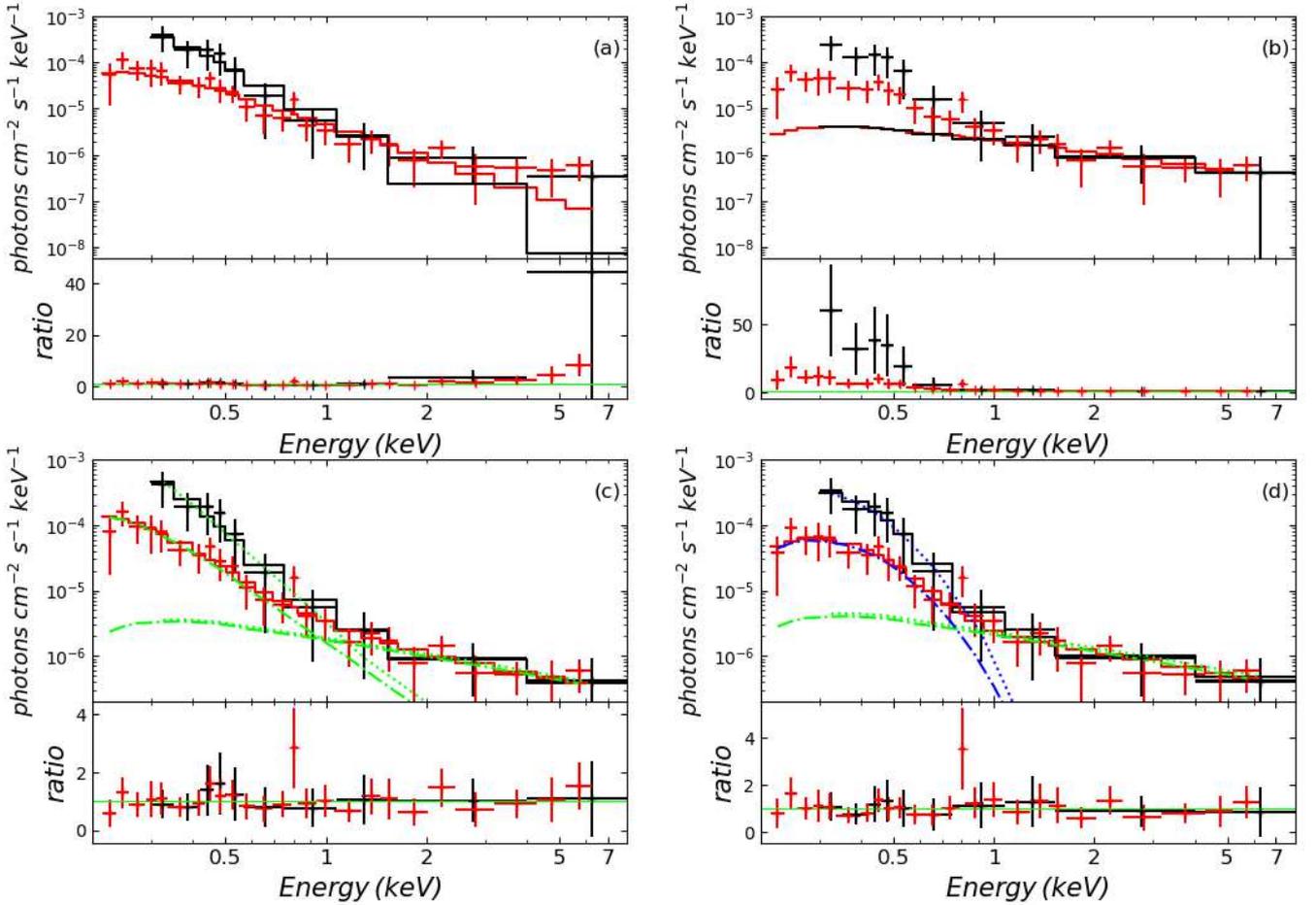}
        \caption{Unfolded {\swift}/XRT and {\xmm}/pn spectra with fitting model. Panel (a): A simple PL model is applied to the full band spectrum of {\swift} (0.3-10\,keV, black) and {\xmm} (0.2-10\,keV, red), respectively. Panel (b): A simple PL model with fixed photon index at $\Gamma=0.9$ is applied to the 1-10\,keV band and extended to the full band spectrum of {\swift} and {\xmm}, respectively. Panel (c):  PL+PL model with fixed photon index at $\Gamma=0.9$ for one PL component is applied to the full band spectrum of {\swift} and {\xmm}, respectively. Panel (d):  PL+BB model with one fixed photon index at $\Gamma=0.9$ for the PL component is  applied to the full band spectrum of {\swift} and {\xmm}, respectively. The data/model ratio is also shown in each panel. }
        \label{spectrum}%
\end{figure*}

First, a simple Galactic absorbed power-law model (PL) with column density fixed at the Galactic value of 1.82$\times$\,10$^{20}$\,cm$^{-2}$ \citep[][the Galactic absorption is included in all following spectral fitting]{kalberla2005leiden} is applied to the {\swift} and {\xmm} spectrum, respectively.  The inferred photon index is $\Gamma=3.9^{+0.8}_{-0.6}$ for {\swift} spectrum (C-statistic/dof=3.8/8)  and $\Gamma=2.4\pm0.3$ for {\xmm} spectrum (C-statistic/dof=19.5/25). However, the residuals in the hard 2-10\,keV band in both of the two spectra  indicate that the other component is needed to well represent the data, as shown in Fig.\ref{spectrum}(a). The PL model is then applied to the {\xmm} 1-10\,keV spectrum because the S/N is better than that of {\swift}. The inferred photon index is $\Gamma=0.9^{+0.4}_{-0.4}$ (C-statistic/dof=2.1/7). We then extrapolated this model to the whole 0.3-10\,keV (0.2-10\,keV) band spectrum of {\swift} ({\xmm}). As shown in Fig.\,\ref{spectrum}(b), there are significant excesses in soft X-ray band, especially for the {\swift} spectrum. 

So we applied the two PL model to the spectra with one photon index fixed at 0.9. The fitting is acceptable and the inferred photon index of the soft PL component is $\Gamma=3.7\pm0.5$ for the {\xmm} spectrum and $\Gamma=4.8^{+1.2}_{-0.9}$ for the {\swift} spectrum.  The soft PL component for the {\swift} spectrum is slightly steeper than that of  {\xmm}, although there is no significant difference considering the uncertainties. But the flux is significantly different (Fig.\,\ref{spectrum}c). The flux in 0.3-2 keV of the soft PL component is $5.11^{+0.91}_{-2.22}\times$10$^{-14}$  and $1.14^{+0.12}_{-0.15}\times$10$^{-14}$ erg\,cm$^{-2}$\,s$^{-1}$,  for {\swift} and {\xmm},  respectively. Notably, there is no flux variation of the hard PL component between the {\swift} and {\xmm} observation.  The flux in 2-10\,keV of the hard PL component is $2.99^{+2.54}_{-1.13}\times10^{-14}$  and 2.80$^{+0.76}_{-0.61}\times$10$^{-14}$\,erg\,cm$^{-2}$ s$^{-1}$ for {\swift} and {\xmm}, respectively.

Since most of the X-ray-selected TDEs can be well fitted with a blackbody model for their soft X-ray band, we used a blackbody (BB) component instead of the soft PL component fitting the spectra (Fig.\,\ref{spectrum}d). The inferred temperature is  $kT=78^{+20}_{-17}$ and $kT=90^{+16}_{-14}$\,eV for {\swift} and {\xmm} spectrum, respectively. The best fitting parameters, including the total fluxes and luminosities in 0.3-2 and 2-10\,keV for each model, are listed in Table\,\ref{table2}.

\begin{table*}
        \caption{Log of the X-ray observations}             
        \label{table1} 
        \centering
        \begin{tabular}{cccccc}
                \hline\hline 
                \makecell[c]{(1)}         &   \makecell[c]{(2)}     &   \makecell[c]{(3)}       &   \makecell[c]{(4)}       & \makecell[c]{(5)}  & \makecell[c]{(6)}  \\
                Tel./instr.   & Obs. ID  &  Obs. date       & Exp. time      & Count rate       & Count rate  \\
                                  &               &                       &                    &   (soft band)      &   (full band)  \\
                                &                 &                       &  [ks]             &  [10$^{-3}$ ct s$^{-1}$] &   [10$^{-3}$ ct s$^{-1}$]  \\
                \hline
                {\swift}/XRT &  00090102001  & 2009 May 15        &  6.73    & 0.80$\pm0.42$  & 1.16$\pm0.52$   \\
                {\swift}/XRT &  00090102002 & 2009 May 17         &  6.60    & 2.38$\pm0.64$  & 2.61$\pm0.69$  \\
                {\swift}/XRT &  00090102003 & 2009 May 20         &  3.26   & 0.56$\pm0.53$  & 0.56$\pm$0.53  \\
                {\swift}/XRT &  stacking         &  2009 May 15-20   &  16.60  & 1.38$\pm$0.32  & 1.54$\pm0.36$   \\
                {\xmm}/pn   & 0843440201    &  2020 January 8    &  12.46  & 9.3$\pm$1.1   & 10.9$\pm$1.2    \\
                \hline
        \end{tabular}
        \tablefoot{--- Col. (1), Telescope and instrument; Col. (2) and Col. (3), Observation ID and date; Col. (4), Effective exposure time; Col (5), Count rate in soft X-ray band (0.3-2\,keV for {\swift}/XRT and 0.2-2\,keV for {\xmm}/pn); Col. (6) Count rate in full X-ray band (0.3-10\,keV for {\swift}/XRT and 0.2-10\,keV for {\xmm}/pn). }
\end{table*}

\begin{table*}
        \caption{X-ray spectral fitting results of {\swift}/XRT and {\xmm}/pn spectra}             
        \label{table2} 
        \centering
        \begin{tabular}{lcccccccc}
                \hline\hline 
                {(1)}   &  {(2)}   &   {(3)}   &  {(4)}   & {(5)}  &  {(6)}   & {(7)}  & {(8)} & {(9)}\\
                Model     & $\Gamma$    &$\Gamma_{1}$  & {kT}       & {C-s/dof}  & ${F_\mathrm{0.3-2\,keV}}$  &${F_\mathrm{2-10\,keV}}$  &  ${L_\mathrm{0.3-2\,keV}}$ & ${L_\mathrm{2-10\,keV}}$ \\
                &         &     &                          &            &   10$^{-14}$   &  10$^{-14}$   & 10$^{41}$  & 10$^{41}$  \\
                &         &     &    {[eV]}              &            &   {[erg cm$^{-2}$ s$^{-1}$]}  &  {[erg cm$^{-2}$ s$^{-1}$]}  & {[erg s$^{-1}$]} & {[erg s$^{-1}$]} \\           
                \hline
                \multicolumn{9}{c}{ {\swift}/XRT (0.3-10\,keV)}  \\
                \hline
                PL        & 3.9$^{+0.8}_{-0.6}$  &                 &   & 3.81/8 & 5.23$^{+1.20}_{-1.46}$ &     0.14$^{+0.01}_{-0.06}$  & 1.74$^{+0.40}_{-0.49}$ & 0.05$^{+0.004}_{-0.020}$ \\
        PL+PL  & 4.8$^{+1.2}_{-0.9}$   & 0.9(fix)   &       & 1.17/6 & 5.65$^{+1.01}_{-2.46}$ & 3.01$^{+2.56}_{-1.14}$ & 1.92$^{+0.38}_{-0.77}$ & 0.90$^{+0.77}_{-0.34}$\\
        PL+BB  &                               &  0.9(fix)  &      78$^{+20}_{-17}$ & 0.75/6 & 5.11$^{+0.81}_{-2.25}$ & 3.71$^{+1.86}_{-1.50}$ & 1.67$^{+0.26}_{-0.74}$ & 1.11$^{+0.56}_{-0.45}$\\
                \hline\hline
                \multicolumn{9}{c}{ {\xmm}/pn (1-10 keV)}  \\
                \hline
            PL  &                     & 0.9$^{+0.4}_{-0.4}$   &   & 2.12/7 &    &     3.32$^{+1.11}_{-1.30}$  &  & 0.94$^{+0.31}_{-0.37}$ \\
            \hline
            \multicolumn{9}{c}{ {\xmm}/pn (0.2-10 keV)}  \\
            \hline
        PL      & 2.4$^{+0.3}_{-0.3}$  &               &     & 19.54/25 & 1.65$^{+0.20}_{-0.17}$ &     0.67$^{+0.13}_{-0.33}$  & 0.52$^{+0.06}_{-0.05}$ & 0.22$^{+0.04}_{-0.11}$ \\
        PL+PL & 3.7$^{+0.5}_{-0.5}$  &0.9(fix) &      & 8.57/22 & 1.64$^{+0.17}_{-0.22}$ & 2.84$^{+0.77}_{-0.62}$ & 0.52$^{+0.05}_{-0.07}$ & 0.85$^{+0.23}_{-0.19}$\\
        PL+BB &                          & 0.9(fix)    &      90$^{+16}_{-14}$ & 9.13/22 & 1.59$^{+0.12}_{-0.23}$ & 3.35$^{+0.66}_{-0.63}$ & 0.50$^{+0.04}_{-0.07}$ & 1.00$^{+0.20}_{-0.19}$\\      
        \hline
                \hline
        \end{tabular}
        \tablefoot{-- Col. (1), Fitting model, PL, PL+PL, and PL+BB are phabs$\mathrm{*}$powerlaw,  phabs$\mathrm{*}$(powerlaw+powerlaw), and phabs$\mathrm{*}$(power law+zbbody), respectively, in Xspec; the column density of $phabs$ is fixed at the Galactic column density of $N_{\mathrm{H}}$=1.82$\times$10$^{20}$\,cm$^{-2}$; col.(2)-(3), Photon index $\Gamma$ of PL component; col. (4), temperature $kT$ of BB component; column(5), best fitting C-statistic/degree of freedom; col. (6)-(7),  in 0.3-2 and 2-10\,keV band after correcting the Galactic absorption; Column(8)-(9), luminosity in 0.3-2 and 2-10\,keV band after correcting the Galactic absorption.}
\end{table*}

\subsection{Radio data reduction}

J1342 was observed at C band (central frequency of 5.5 GHz) with the VLA in its moderately compact C configuration on March 12, 2016 (program code, 15B-247; PI: Zauderer). The data were reduced following standard procedures with the CASA package. Flux density calibration was conducted using 3C286, whereas the nearby source J1347+1217 was used to determine the complex gain solutions, which were interpolated to J1342. After removing the radio frequency interference, the data were imaged using the CLEAN algorithm with Briggs weighting and ROBUST parameter of 0. The final cleaned map has a synthesized beam of 3.2\arcsec$\times$2.4\arcsec. J1342 is clearly detected as a compact source, with an integrated flux density of $44.9\pm2.1\,\mu$Jy, which was measured using the CASA task IMFIT.


\section{Discussion} \label{3}

\subsection{X-ray origin}
The two X-ray spectra can be well described with a two-component model: one component is a hard PL component with photon index of $\sim1.0$, dominated in the 2-10\,keV band; the other is a very soft component as a steep PL with photon index of $\Gamma=4.8^{+1.2}_{-0.9}$ for {\swift} (or $\Gamma=$3.7$\pm$0.5 for {\xmm}) or a blackbody with temperature of $kT\sim70-90$\,{eV}, dominated in the 0.2-2\,keV band. In the {\swift} observation, the luminosity of the hard component is  $\sim$1.6 and 8.9$\times10^{40}$\,{erg\,s$^{-1}$} in 0.3-2 and 2-10\,keV, respectively; the soft component is  $\sim$1.8  and 0.008$\times10^{41}$\,{erg\,s$^{-1}$} in 0.3-2 and 2-10\,keV, respectively. In the {\xmm} observation, the luminosity of the hard component is  $\sim1.5$  and 8.4$\times10^{40}$\,{erg\,s$^{-1}$} in 0.3-2 and 2-10\,keV, respectively; the soft component is  $\sim$3.7 and $0.16\times10^{40}$\,{erg\,s$^{-1}$} in 0.3-2 and 2-10\,keV, respectively. 


The host galaxy of J1342 is a star-forming galaxy of which the hot interstellar medium could contribute to X-ray emission. The estimated X-ray emission from its host galaxy is $\sim5\times10^{38}$\,{erg\,s$^{-1}$} in 0.3-10\,keV using the correlation between the star formation rate (SFR) and the X-ray luminosity \citep[e.g.,][]{mineo2012x}, if adopting its SFR of 0.07\,${\it M}$$_{\sun}$\,yr$^{-1}$\,\citep[][]{french2020host}. This is more than two orders of magnitude lower than the luminosity observed by {\swift} or {\xmm}. Thus, the host galaxy origin for the observed X-ray emission could be ruled out. 

The 0.3-2\,keV band luminosity of {\swift} observation is $\sim3.7$ times brighter than that of {\xmm}. If only considering the soft PL component, the luminosity of {\swift} would be $\sim4.9$ times brighter than that of {\xmm}. However, there is no significant difference in the 2-10\,keV band luminosity between {\swift} and {\xmm}. This indicates that the emissions of the soft and hard component may have different origins for J1342. 

As the luminosity of the hard component in 0.3-10\,keV is $1.0-1.1\times10^{41}$\,{erg\,s$^{-1}$} and no significant variation is detected in between the {\swift} and {\xmm} observations, we speculate this component has always existed. The origin of this hard component is unclear. One possibility is that J1342 has some low level of  AGN activity. Using the luminosity correlation between ${L_{[\mathrm{O}_{\,\mathrm{III}}]}}$ and $L_{X}$ in AGNs \citep[e.g.,][]{Lamastra2009}, we can also estimate the X-ray contribution of AGN activity from its  ${{[\mathrm{O}_{\, \mathrm{III}}]}}$ narrow emission line. Adopting the ${{[\mathrm{O}_{\, \mathrm{III}}]}}$ flux of 1.8$\times10^{-15}$\,{erg\,cm$^{-2}\,$s$^{-1}$} \citep{yang2013long}, the estimated 2-10\,keV luminosity is $7.5\times10^{40}$\,{erg\,s$^{-1}$}, which is consistent with the observed 2-10\,keV luminosity of the hard component. This probability is similar to the speculation of \citet{chilingarian2018population}.

Soft X-ray excess is a common feature in some AGNs, which is also usually modeled with a blackbody or a steep power law. However, such a steep soft X-ray spectrum is  usually detected in X-ray bright AGNs, especially, in NLS1ys \citep[e.g.,][]{ai2010x,gliozzi2020soft}.  Additionally, the variability of soft excess is generally concurrent with its disk-coronal emissions in the hard X-ray band. For J1342, the variability is only detected in the soft X-ray band and the observed luminosity is relatively lower than such bright AGNs.

\begin{figure*}
        \centering
        \includegraphics[width=0.9\textwidth]{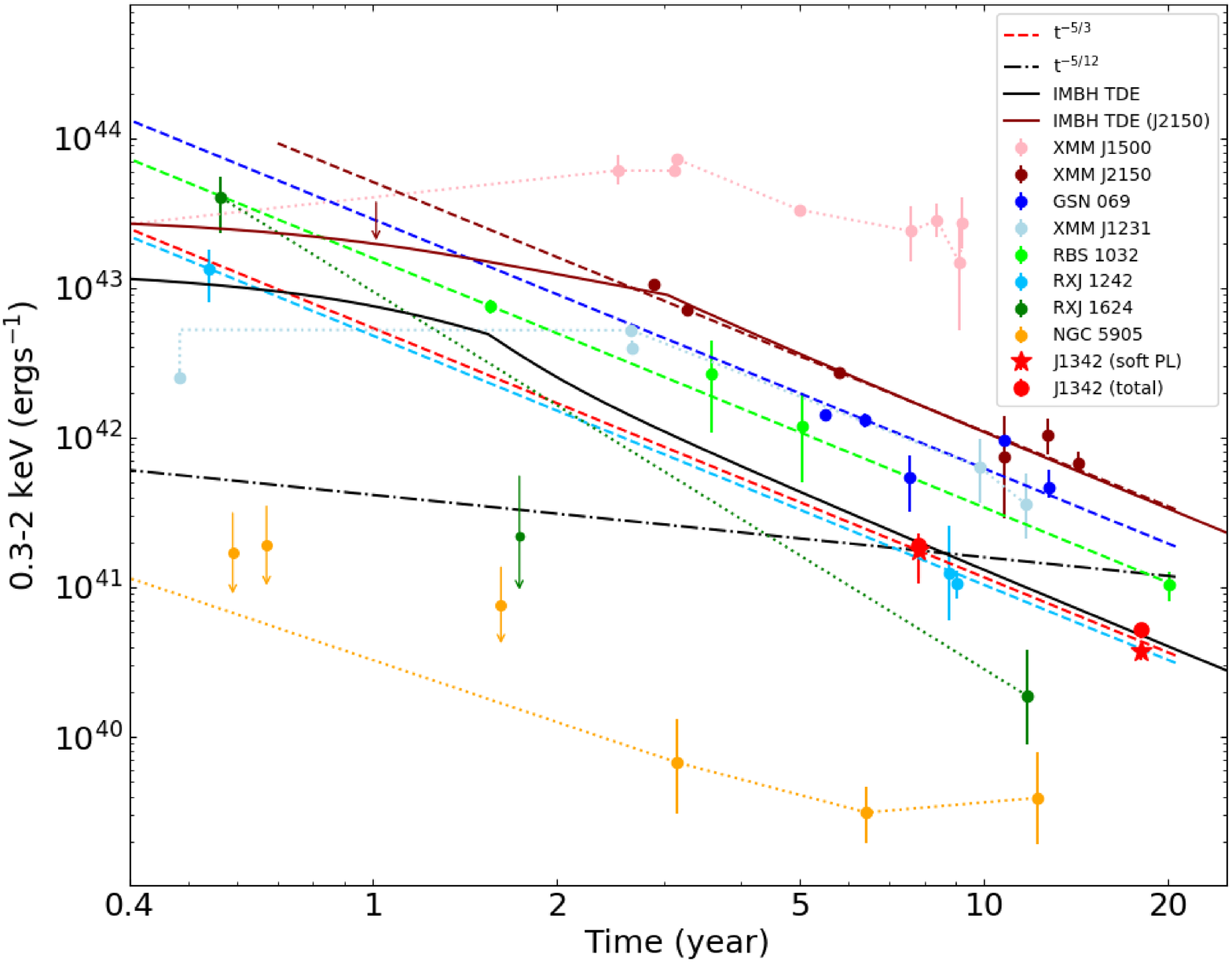}
        \caption{Later X-ray light curves of TDEs in 0.3-2\,keV. The x-axis indicates the time since the flare occurred; for J1342, it is about 1 year before the SDSS optical spectral observation. The total 0.3-2\,keV luminosities of J1342 is labeled as a circle symbol; the 0.3-2\,keV luminosities of the soft PL component of J1342 is labeled as a pentagram symbol. The later X-ray light curves of other long-lived TDEs are also shown, which include RX J1242.6-1119, RX J1624.9+7554, NGC5905\,\citep{vaughan2004chandra}, RBS 1032\,\citep{maksym2014rbs}, 3XMM J1500+0154\,\citep{Lin2017a}, 3XMM J2150-0551\,\citep{Lin2018,Lin2020}, 2XMM J1231+1106\,\citep{Lin2017b}, and GSN 069\,\citep{Shu2018}. The luminosities of XMM J1500+0154 are in  0.34-11.5\,keV, of RBS 1032 are in 0.1-2.4\,keV,  and of XMM J2150-0551 are the bolometric luminosities from the disk model; the luminosities of others are converted to those in the 0.3-2\,keV band. The later light curves following the power law decline as ${{\it t}^{-5/3}}$ are shown as dashed lines; the others are shown as dotted lines.  A power law decline as ${{\it t}^{-5/12}}$ is also shown as a dot-dashed line. The theoretical IMBH-TDE light curve\,\citep{Chen2018} is shown as a solid line.}
        \label{xray-lc}
\end{figure*}

As a TDE candidate with transient coronal lines, variability in the soft X-ray band is expected in J1342. The light curve of 0.3-2\,keV band luminosity is shown in Fig.\,\ref{xray-lc}. We apply a declined power-law function to fit the light curve as follows:
$${L_\mathrm{0.3-2\,keV}}\,=\,{A\,{\it t}^{-n}},$$
where $t$ is time since the date of the X-ray peak of the tidal disruption flare in unit of year. Here, we adopt the peak of the X-ray flare occurred about one year prior to the SDSS optical spectral observation as estimated in \citet{dou2016long}. If we adopt the total X-ray luminosities in 0.3-2\,keV band, we obtain a power-law index of $n\,=\,1.6^{+0.2}_{-0.5}$. Fitting to the X-ray luminosity evolution of the soft component in 0.3-2\,keV band, we obtain a power-law index of $n\,=\,1.8^{+0.3}_{-0.5}$. 
Interestingly, we find the observed X-ray luminosities in 0.3-2 keV well matched the theoretical canonical decline of ${{\it t}^{-5/3}}$ in the fallback model \citep{rees1988tidal,Phinney1989} within errors, although X-ray observations are 8 ({\swift}) and 19 ({\xmm}) years after the outburst (as shown in Figure\,\ref{xray-lc}). In constrat, ${{\it t}^{-5/12}}$ declined power law, predicted with the assumption that the X-ray emissions are from the disk of TDE \citep[e.g.,][]{cannizzo2009a,lodato2011multiband}, deviates from the data. Furthermore, the expected 0.3-2\,keV band luminosity peak is at $\sim10^{44}$\,{erg\,s$^{-1}$}, which is also consistent with the luminosity peak in other soft X-ray discovered TDEs \citep{komossa2015tidal,saxton2020x}. Additionally, the spectral slope of the soft component of J1342 is $\Gamma=4.8^{+1.2}_{-0.9}$ for {\swift} and $\Gamma=3.7\pm0.5$ for {\xmm}, or  $kT=78^{+20}_{-17}$\,{eV} for {\swift}, and $kT=90^{+16}_{-14}$\,{eV} for {\xmm} if modeling with a blackbody. 

\subsection{Radio origin}
  J1342 is clearly detected as a compact radio source, with an integrated flux density of $44.9\pm2.1\,\mu$Jy at 5.5\,GHz. There are three possible origins for the radio emission: star formation, AGN, and/or TDE. If assuming the radio-derived SFR ({$\mathrm{SFR}_{\mathrm{radio}}$}) equal to the optical SFR of 0.07\,${\it M}_{\sun}$\,yr$^{-1}$, we can estimate the radio contribution from star formation, using the relation between the SFR  and radio flux \citep[e.g., the expression in Section 3.2 in][]{Greiner2016}. After subtracting the star formation contribution, the radio emission from the AGN and/or TDE is obtained at a level of $\sim$31\,${\mu}$Jy. As discussed in Section 3.1, a persistent harder component from AGN and a variable soft component from TDE are present in the X-ray spectra of J1342. Since only a single-epoch radio observation is available, we cannot tell whether there is a similar radio variability as observed in soft X-rays, which is helpful to identify the radio origin from a TDE. Further observations at radio bands are encouraged to determine whether the radio emission is stable or declining in flux. 
  
 \subsection{ Black hole mass} 

The BH mass of J1342 is estimated to be $1.1\times10^{6}$\,${\it M}_{\sun}$ with an uncertainty of 0.5\,dex using the ${\mathrm M-\sigma}$ correlation \citep{french2020host}. However, \citet{chilingarian2018population} measured its virial BH mass as $9.6\pm1.3\times10^{4}$\,${\it M}_{\sun}$ using the flux and width of the broad H$\alpha$ emission from the Magellan/MagE optical spectrum taken on May 30, 2017, which is $\sim16$ years after the flare. Assuming the host-subtracted radio emission originates from a persistent AGN, we can also estimate the BH mass, using the fundamental plane of BH accretion, an empirical correlation of the mass of a BH, its radio luminosity, and its 2-10\,$\mathrm{keV}$ band power-law continuum luminosity \citep[][]{Gultekin2014}. The BH mass estimated from this correlation is 4.3$\times$10$^{5}$\,${\it M}_{\sun}$, with an uncertainty of 0.8\,dex. Besides the intrinsic scatter, the low luminosity of J1342 makes the correction for radio contamination from host emission uncertain. Therefore, BH mass estimation from the fundamental plane should be treated with caution.

Considering the scatter of ${\mathrm M-\sigma}$ correlation at low BH mass ($<10^{6}$\,{${\it M}_{\sun}$}) is very large \citep{Kormendy2013}, the estimated BH mass of J1342 based on this correlation may have large intrinsic uncertainty. The estimation from a fundamental plane correlation has an even large scatter and whether such a correlation is applicable to SMBHs is still under debate \citep{Gultekin2019}. The systematic uncertainty of the broad H$\alpha$-based virial estimation for the BH mass is usually within 0.3\,dex \citep{Xiao2011, Dong2012}. Furthermore, the Magellan/MagE spectrum is obtained 16 years after the flare, and it would scarcely be affected by the flare. Based on these reasons, we prefer the BH mass of $9.6\times10^{4}$\,${\it M}$$_{\sun}$ from the broad H$\alpha$-based virial estimation, and we consider  the values obtained from the other two methods only for a consistency check;  these checks support the presence of an intermediate mass BH (IMBH; $10^{2}$\,{${\it M}_{\sun}\,<$}\,BH mass\,$<\,10^{5}$\,${\it M}_{\sun}$) in J1342.

\subsection{IMBH-TDE}

The X-ray spectral slope and long-term spectral evolution is similar to other thermal X-ray TDEs, for example, RBS 1032. The  luminosity decline of RBS 1032 also follows the ${{\it t}^{-5/3}}$ approximation and has  a super-soft spectrum (${\Gamma\sim5}$) at its early phase and remains still very soft ($\Gamma\sim3.4$) 20 years after the discovery \citep{maksym2014rbs}. The similar X-ray evolution behavior of J1342 suggests it is also a TDE, which is consistent with the previous studies from its transient coronal lines (Wang et al. 2012; Yang et al. 2013) and mid-infrared dust echo (Dou et al. 2016). If so, it would make J1342 one of the long-lived TDEs lasting more than ten years.

The long-lived TDEs have been discovered in several individuals, including RX J1242.6-1119, RX J1624.9+7554, NGC5905 \citep{vaughan2004chandra}, RBS 1032 \citep{maksym2014rbs}, 3XMM J1500+0154 \citep{Lin2017a}, 3XMM J2150-0551 \citep{Lin2018,Lin2020}, 2XMM J1231+1106 \citep{Lin2017b}, and GSN 069 \citep{Shu2018}, although the later two source is a TDE occurring in an AGN. We collected their X-ray light curves and plotted these in Figure\,\ref{xray-lc}. We note that the ${\it t}^{-5/3}$ declines in their very later phase are also discovered in XMM J2150-0551, GSN069, RX J1242.6-1119, and RBS 1032.

Several theoretical models could explain the long-lived TDEs, for example, a long super-Eddington accretion phase, perhaps involving a disrupted object with a large mass \citep[e.g.,][]{Lin2017a}, a partially stripped evolved star atmosphere \citep{MacLeod2012}, or later distant circularization \citep[][see also Saxton et al. 2020]{Guillochon2015,Shiokawa2015}. For J1342, if adopted the virial BH mass of $\sim9.6\times10^{4}$\,${\it M}$$_{\sun}$, the X-ray luminosity peak of J1342 estimated from the power-law decline of ${{\it t}^{-5/3}}$ would be one magnitude order larger than its Eddington luminosity (L$_{\mathrm{Edd}}$\,=\,1.26\,$\times$\,10$^{38}$\,(${\mathrm M}_{\mathrm {BH}}$/${\it M}_{\sun}$)\,erg\,s$^{-1}$). Additionally, these would also have a super-Eddington stage lasting roughly one year in the early stage of outburst.

\citet{Chen2018} suggest an IMBH-TDE model explains the more than 13 year long-lived TDE 3XMM J2150-0551. These authors predicted that the tidal disruption of a main-sequence star by an IMBH could always be a long-term circularization process, leading a light curve as long as two decades, which still follows the ${\it t}^{-5/3}$ decline in a very later phase. We can also explain the later-time X-ray evolution of J1342 with this IMBH-TDE model, as shown in Figure\,\ref{xray-lc}, for example, the tidal disruption of a main-sequence star with mass of 0.1\,${\it M}_{\sun}$, adopting the virial BH mass of $\sim9.6\times10^{4}\,{\it M}_{\sun}$.

\section{Conclusion} \label{4}

X-ray emission was detected in the TDE J1342 from {\swift}/{XRT} and {\xmm}/{pn} observations, which is $\sim9$ and 19 years after the flare. The luminosity of the {\xmm} observation is fainter at ${L_{0.3-10\,\mathrm{keV}}}$\,$\sim\,1.4\times10^{41}$\,{erg\,s$^{-1}$}, compared to the value, ${L_{0.3-10\,\mathrm{keV}}}$\,$\sim\,2.8\times10^{41}$\,{erg\,s$^{-1}$} of {\swift}. We find, however, that the variation is mostly from the emission in soft X-ray:  ${L_{0.3-2\,\mathrm{keV}}}$\,$\sim\,5.2\times10^{40}$\,{erg\,s$^{-1}$} in {\xmm} observation and ${L_{0.3-2\,\mathrm{keV}}}$\,$\sim1.9\times10^{41}$\,{erg\,s$^{-1}$} in {\swift}. The emission in hard 2-10\,keV band is consistent between the two observations.

To infer the nature of the variation we modeled the spectra with two-component models, including one flat hard power law plus one steep power law, or a blackbody component. The flat, hard power-law component could be explained as the persistent AGN activity. We find some TDE evidence as follows:

\begin{itemize}
\item The soft component can be well fitted with a steep power law or blackbody model. The spectra is somewhat harder at fading phase of duration  years with $\Gamma=3.7\pm0.5$ as obtained from {\xmm}, compared to  $\Gamma=4.8^{+1.2}_{-0.9}$ as obtained with {\swift}, although the spectral slopes are consistent with each other if we consider the uncertainties. In a blackbody modeling  the inferred temperature is $kT=78^{+20}_{-17}$\,{eV}, and $kT=90^{+16}_{-14}$\,{eV} in the spectra of {\swift} and {\xmm}, separately.
\item The 0.3-2\,keV band luminosity of the soft component fades from 1.8$\times10^{41}$ to 3.7$\times10^{40}$\,{erg\,s$^{-1}$} in $\sim10$ years. The declination of the luminosity is well fitted with ${\it t}^{-5/3}$ model.
\item The estimated peak luminosity of the flare is  $\sim10^{44}$\,{erg\,s$^{-1}$} with an assumption of ${\it t}^{-5/3}$ declination.
\end{itemize}

Therefore, the soft X-ray spectral shape and luminosity evolution further support that J1342 is a TDE candidate, which justifies the previous studies with transient coronal lines (Wang et al. 2012; Yang et al. 2013) and mid-infrared dust echo (Dou et al. 2016). Furthermore, if the BH mass of $\sim9.6\times10^4\,{\it M}_{\sun}$ estimated from the H$\alpha$ emission in the 16 year later-time optical spectrum is reliable, J1342 would be a long-lived IMBH-TDE candidate.

\begin{acknowledgements}

We would like to thank the anonymous referee and the editor Prof. Sergio Campana for the suggestions and comments that greatly improved the paper. This work is supported by Joint Research Foundation in Astronomy (U1731104, U1731109, U2031106) under cooperative agreement between the NSFC and the CAS, Chinese Science Foundation (NSFC-11833007, 11822301, 11733001). L.M.D. and J.S.H. also acknowledge the support from the Key Laboratory for Astronomical Observation and Technology of Guangzhou, the Astronomy Science and Technology Research Laboratory of Department of Education of Guangdong Province, the opening fund of CAS Key Laboratory of Galaxy Cosmology (No.18010201). 

\end{acknowledgements}


\begin{thebibliography}{}
        \expandafter\ifx\csname natexlab\endcsname\relax\def\natexlab#1{#1}\fi
        
        \bibitem[{Ai {et~al.}(2010)Ai, Yuan, Zhou, Wang, Zhang}]{ai2010x} Ai, Y., Yuan, W., Zhou, H., Wang, T., \& Zhang, S. 2010, \apj, 727, 31
        
        \bibitem[{Auchettl {et~al.}(2017)Auchettl, Guillochon, Ramirez-Ruiz}]{auchettl2017new} Auchettl, K., Guillochon, J., \& Ramirez-Ruiz, E. 2017, \apj, 838, 149

        \bibitem[{Bade {et~al.}(1996)Bade, Komossa, Dahlem}]{bade1996detection} Bade, N., Komossa, S., \& Dahlem, M. 1996, \aap, 309, L35
        
        
        \bibitem[{Burrows {et~al.}(2005)Burrows, Hill, Nousek, Kennea, Wells, Osborne, Abbey, Beardmore, Mukerjee, Short, {et~al.}}]{burrows2005swift}        Burrows, D.~N., Hill, J., Nousek, J.~A., {et~al.} 2005, \ssr, 120, 165
        
        
        \bibitem[{Cannizzo \& Gehrels(2009)}]{cannizzo2009a} Cannizzo, J. \& Gehrels, N. 2009, \apj, 700, 1047

        \bibitem[Chen \& Shen(2018)]{Chen2018} Chen, J.-H. \& Shen, R.-F.\ 2018, \apj, 867, 20 
                
        \bibitem[{Chilingarian {et~al.}(2018)Chilingarian, Katkov, Zolotukhin, Grishin, Beletsky, Boutsia, \& Osip}]{chilingarian2018population} Chilingarian, I.~V., Katkov, I.~Y., Zolotukhin, I.~Y., {et~al.} 2018, \apj, 863, 1
        
       \bibitem[Dong et al.(2012)]{Dong2012} Dong, X.-B., Ho, L.~C., Yuan, W., et al.\ 2012, \apj, 755, 167   

        \bibitem[{Dou {et~al.}(2016)Dou, Wang, Jiang, Yang, Lyu, \& Zhou}]{dou2016long} Dou, L., Wang, T.-g., Jiang, N., {et~al.} 2016, \apj, 832, 188
        
        \bibitem[{Dou {et~al.}(2017)Dou, Wang, Yan, Jiang, Yang, Cutri, Mainzer, \& Peng}]{dou2017discovery}
        Dou, L., Wang, T., Yan, L., {et~al.} 2017, \apjl, 841, L8
        
        \bibitem[{Evans \& Kochanek(1989)}]{evans1989the} Evans, C.~R. \& Kochanek, C.~S. 1989, \apj, 346

        \bibitem[{French {et~al.}(2020)French, Wevers, Law-Smith, Graur, \& Zabludoff}] {french2020host} French, K.~D., Wevers, T., Law-Smith, J., Graur, O., \& Zabludoff, A.~I. 2020, \ssr, 216, 1       
        
        \bibitem[{Gehrels {et~al.}(2004)Gehrels, Chincarini, Giommi, Mason, Nousek, Wells, White, Barthelmy, Burrows, Cominsky, {et~al.}}]{gehrels2004swift} Gehrels, N., Chincarini, G., Giommi, P., {et~al.} 2004, \apj, 611, 1005
        
                
        \bibitem[{Gliozzi \& Williams(2020)}]{gliozzi2020soft} Gliozzi, M. \& Williams, J.~K. 2020, \mnras, 491, 532
        
        \bibitem[Greiner et al.(2016)]{Greiner2016} Greiner, J., Micha{\l}owski, M.~J., Klose, S., et al.\ 2016, \aap, 593, A17 


       \bibitem[Guillochon \& Ramirez-Ruiz(2015)]{Guillochon2015} Guillochon, J. \& Ramirez-Ruiz, E.\ 2015, \apj, 809, 166 
        
        \bibitem[G{\"u}ltekin et al.(2014)]{Gultekin2014} G{\"u}ltekin, K., Cackett, E.~M., King, A.~L., et al.\ 2014, \apjl, 788, L22 
        
        \bibitem[G{\"u}ltekin et al.(2019)]{Gultekin2019} G{\"u}ltekin, K., King, A.~L., Cackett, E.~M., et al.\ 2019, \apj, 871, 80 
        

        \bibitem[{Hills(1975)}]{hills1975possible} Hills, J.~G. 1975, \nat, 254, 295
        
        \bibitem[{Jiang {et~al.}(2016)Jiang, Dou, Wang, Yang, Lyu, \& Zhou}]{jiang2016wise} Jiang, N., Dou, L., Wang, T., {et~al.} 2016, \apjl, 828, L14    
                
        \bibitem[{Jiang {et~al.}(2017) Jiang, Wang, Yan, Xiao, Yang, Dou, Wang, Cutri, \& Mainzer}]{jiang2017mid} Jiang, N., Wang, T., Yan, L., {et~al.} 2017, \apj, 850, 63
        
        \bibitem[{Jiang {et~al.}(2019) Jiang, Wang, Mou, Liu, Dou, Sheng, \& Wang}]{jiang2019infrared} Jiang, N., Wang, T., Mou, G., {et~al.} 2019, \apj, 871, 15   
        
        \bibitem[{Jiang {et~al.} (2021) Jiang, Wang, Dou, Shu, Hu, Liu, Wang, Yan, Sheng, Yang, Sun, Zhou}] {jiang2021} Jiang, N., Wang, T., Dou, L., {et~al.}\ 2021, \apjs, 252, 32. 
                        
        \bibitem[{Jonker {et~al.}(2020)Jonker, Stone, Generozov, van Velzen, \& Metzger}] {jonker2020implications} Jonker, P., Stone, N., Generozov, A., van Velzen, S., \& Metzger, B. 2020, \apj, 889, 166
        
        \bibitem[{Kalberla {et~al.}(2005)Kalberla, Burton, Hartmann, Arnal, Bajaja, Morras, \& P{\"o}ppel}]{kalberla2005leiden} Kalberla, P.~M., Burton, W., Hartmann, D., {et~al.} 2005,\aap , 440, 775
        
        \bibitem[{Komossa \& Bade(1999)}]{komossa1999the} Komossa, S. \& Bade, N. 1999,\aap, 343, 775
                
        \bibitem[{Komossa {et~al.}(2008)Komossa, Zhou, Wang, Ajello, Ge, Greiner, Lu, Salvato, Saxton, Shan, {et~al.}}]{komossa2008discovery} Komossa, S., Zhou, H., Wang, T.~G., {et~al.} 2008, \apj, 678
        
        \bibitem[{Komossa(2015)}]{komossa2015tidal} Komossa, S. 2015, Journal of High Energy Astrophysics, 7, 148
                
        \bibitem[Kormendy \& Ho(2013)]{Kormendy2013} Kormendy, J. \& Ho, L.~C.\ 2013, \araa, 51, 511  

        
        \bibitem[Lamastra et al.(2009)]{Lamastra2009} Lamastra, A., Bianchi, S., Matt, G., et al.\ 2009, \aap, 504, 73.
                
        \bibitem[{Lin {et~al.}(2017{\natexlab{a}})Lin, Godet, Ho, Barret, Webb, \&
                        Irwin}]{Lin2017a}
        Lin, D., Godet, O., Ho, L.~C., {et~al.} 2017{\natexlab{a}}, \mnras, 468, 783
        
        \bibitem[{Lin {et~al.}(2017{\natexlab{b}})Lin, Guillochon, Komossa,
                Ramirezruiz, Irwin, Maksym, Grupe, Godet, Webb, Barret, {et~al.}}]{Lin2017b}
        Lin, D., Guillochon, J., Komossa, S., {et~al.} 2017{\natexlab{b}}, Nature
        Astronomy, 1, 0033      
        
        \bibitem[Lin et al.(2018)]{Lin2018} Lin, D., Strader, J., Carrasco, E.~R., et al.\ 2018, Nature Astronomy, 2, 656
        
        \bibitem[Lin et al.(2020)]{Lin2020} Lin, D., Strader, J., Romanowsky, A.~J., et al.\ 2020, \apjl, 892, L25.
                        
        \bibitem[{Lu {et~al.}(2016)Lu, Kumar, \& Evans}]{lu2016infrared} Lu, W., Kumar, P., \& Evans, N.~J. 2016, \mnras, 458, 575
        
        \bibitem[{Lodato \& Rossi(2011)}]{lodato2011multiband}
        Lodato, G. \& Rossi, E.~M. 2011, \mnras, 410, 359
        
        \bibitem[MacLeod et al.(2012)]{MacLeod2012} MacLeod, M., Guillochon, J., \& Ramirez-Ruiz, E.\ 2012, \apj, 757, 134 
        
        \bibitem[{Maksym {et~al.}(2014)Maksym, Lin, \& Irwin}]{maksym2014rbs} Maksym, W.~P., Lin, D., \& Irwin, J.~A. 2014, \apj, 792, L29
        
        \bibitem[{Mineo {et~al.}(2012)Mineo, Gilfanov, \& Sunyaev}]{mineo2012x} Mineo, S., Gilfanov, M., \& Sunyaev, R. 2012,\mnras, 426, 1870
        
        
        \bibitem[{Park {et~al.}(2006)Park, Kashyap, Siemiginowska, Van~Dyk, Zezas, Heinke, \& Wargelin}]{park2006bayesian} Park, T., Kashyap, V., Siemiginowska, A., {et~al.} 2006, \apj, 652, 610

        \bibitem[{Phinney(1989)Phinney E. S. }]{Phinney1989} Phinney E. S., 1989, in Morris M., ed., IAU Symp. Vol. 136, Manifestationsof  a  Massive  Black  Hole  in  the  Galactic  Center.  Kluwer,  Dordrecht, p. 543    
        
        \bibitem[{Rees(1988)}]{rees1988tidal} Rees, M.~J. 1988, \nat, 333, 523     

        
        \bibitem[{Saxton {et~al.}(2020)Saxton, Komossa, Auchettl, \& Jonker}]{saxton2020x} Saxton, R., Komossa, S., Auchettl, K., \& Jonker, P. 2020, \ssr, 216, 1
                
        \bibitem[Shiokawa et al.(2015)]{Shiokawa2015} Shiokawa, H., Krolik, J.~H., Cheng, R.~M., et al.\ 2015, \apj, 804, 85.       
        
        \bibitem[{Shu {et~al.}(2018)Shu, Wang, Dou, Jiang, Wang, \& Wang}]{Shu2018} Shu, X., Wang, S.~S., Dou, L., {et~al.} 2018, \apjl, 857
        
        
        \bibitem[{Vaughan {et~al.}(2004)Vaughan, Edelson, \& Warwick}]{vaughan2004chandra} Vaughan, S., Edelson, R., \& Warwick, R. 2004, \mnras, 349, L1
        

        \bibitem[{van Velzen {et~al.}(2016)van Velzen, Mendez, Krolik, \& Gorjian}]{van2016discovery} van Velzen, S., Mendez, A.~J., Krolik, J.~H., \& Gorjian, V. 2016, \apj, 829, 19
        
        \bibitem[{Wang {et~al.}(2011)Wang, Zhou, Wang, Lu, \& Xu}]{wang2011transient} Wang, T.-G., Zhou, H.-Y., Wang, L.-F., Lu, H.-L., \& Xu, D. 2011, \apj, 740, 85
        
        \bibitem[{Wang {et~al.}(2012)Wang, Zhou, Komossa, Wang, Yuan, \& Yang}]{wang2012extreme} Wang, T.-G., Zhou, H.-Y., Komossa, S., {et~al.} 2012, \apj, 749, 115
        
        \bibitem[{Wang {et~al.}(2018)Wang, Yan, Dou, Jiang, Sheng, \& Yang}]{wang2018long} Wang, T., Yan, L., Dou, L., {et~al.} 2018, \mnras, 477, 2943
        
        \bibitem[Xiao et al.(2011)]{Xiao2011} Xiao, T., Barth, A.~J., Greene, J.~E., et al.\ 2011, \apj, 739, 28   
                        
        \bibitem[{Yang {et~al.}(2013)Yang, Wang, Ferland, Yuan, Zhou, \& Jiang}]{yang2013long} Yang, C.-W., Wang, T.-G., Ferland, G., {et~al.} 2013, \apj, 774, 46   
        
        
        
\end{thebibliography}
\end{document}